\documentclass[sigconf,screen]{acmart}
\usepackage{subcaption}
\usepackage{algorithm}
\usepackage{balance}
\usepackage{microtype}
\usepackage{algpseudocode}
\usepackage{float}
\usepackage{multirow}
\usepackage{booktabs}
\usepackage{url}
\usepackage{tcolorbox}
\usepackage{xcolor}

\newtcolorbox{InsightBox}[1]{
  colback=white,          
  colframe=black,         
  colbacktitle=black,     
  coltitle=white,         
  fonttitle=\bfseries,    
  title={#1},             
  sharp corners,          
  boxrule=0.7pt,          
  left=4pt, right=4pt, top=4pt, bottom=4pt, 
  after skip=10pt         
}

\setcitestyle{nocompress, sort}

\hypersetup{
    colorlinks=true,   
    citecolor=blue,    
    linkcolor=blue,    
    urlcolor=blue      
}

\AtBeginDocument{%
  }

\setcopyright{cc}
\setcctype{by}
\acmDOI{10.1145/3832783.3837458}
\acmYear{2026}
\copyrightyear{2026}
\acmISBN{979-8-4007-2882-2/2026/10}
\acmConference[ASE '26]{Proceedings of the 41st IEEE/ACM International Conference on Automated Software Engineering}{October 12--16, 2026}{Munich, Germany}
\acmBooktitle{Proceedings of the 41st IEEE/ACM International Conference on Automated Software Engineering (ASE '26), October 12--16, 2026, Munich, Germany}
\acmSubmissionID{ase26main-p732-p}
\received{2026-03-26}
\received[accepted]{2026-06-18}

\begin{document}

\title{From Guessing to Seeing: Enhancing LLM-Based Program Repair via Trace-Guided Multi-strategy Debate}


\author{Jiaqing Wu}
\orcid{0009-0000-0730-8313}
\affiliation{%
  \institution{Northwestern Polytechnical University}
  \city{Xi'an}
  \country{China}
}
\email{jiaqingwu@mail.nwpu.edu.cn}

\author{Tong Wu}
\orcid{0009-0008-6675-2201}
\affiliation{%
  \institution{Northwestern Polytechnical University}
  \city{Xi'an}
  \country{China}
}
\email{wutong2003@mail.nwpu.edu.cn}

\author{Manqing Zhang}
\orcid{0000-0001-9086-0503}
\affiliation{%
  \institution{Northwestern Polytechnical University}
  \city{Xi'an}
  \country{China}
}
\email{zmqgeek@mail.nwpu.edu.cn}

\author{Yunwei Dong}
\orcid{0000-0001-9882-9121}
\affiliation{%
  \institution{Northwestern Polytechnical University}
  \city{Xi'an}
  \country{China}
}
\email{yunweidong@nwpu.edu.cn}

\author{Bo Shen}
\correspondingauthor
\orcid{0000-0002-8175-3789}
\affiliation{%
  \institution{Northwestern Polytechnical University}
  \city{Xi'an}
  \country{China}
}
\email{shen@nwpu.edu.cn}
\renewcommand{\shortauthors}{Jiaqing Wu, Tong Wu, Manqing Zhang, Yunwei Dong, and Bo Shen}

\begin{abstract}
Automated Program Repair (APR) aims to resolve software bugs without human intervention. However, handling complex logic errors, especially silent failures that produce incorrect outputs without any explicit crash signals, remains a significant challenge. Current LLM-based APR approaches attempt to address this issue through conversational feedback, retrieval-augmented generation, or static analysis tools. However, these methods are fundamentally static, relying only on source code and basic test outputs, and thus struggle to capture complex runtime behaviors and dynamic data dependencies accurately. Inspired by human debugging practice, incorporating runtime evidence into APR provides direct access to program behavior through execution traces, which expose concrete state transitions and data dependencies. However, even with such runtime evidence, a single LLM interpreting it in isolation tends to commit to a specific repair hypothesis, leading to test overfitting and producing patches that satisfy the observed traces and test suite by coincidence rather than correct logic. This observation suggests that runtime evidence should not be treated merely as additional input, but as objective constraints that candidate patches must satisfy. Building on this insight, we propose \textsc{TraceRepair}, a multi-agent framework that leverages runtime facts as shared constraints for patch validation. A probe agent captures execution snapshots of critical variables, forming an objective basis for repair, while a committee of specialized agents cross-verifies candidate patches to expose inconsistencies and iteratively refine them. Evaluated on the Defects4J benchmark, \textsc{TraceRepair} substantially improves repair effectiveness, correctly fixing 392 defects and outperforming existing LLM-based approaches. Extensive experiments further demonstrate improved efficiency and strong generalization on a newly constructed dataset of recent bugs, suggesting that the performance gains arise from dynamic reasoning rather than memorization.
\end{abstract}




\maketitle

\section{Introduction}
\label{sec:intro}

Automated Program Repair (APR) aims to automatically transform buggy programs into correct ones without human intervention, thereby mitigating the high cost of manual debugging~\cite{le2011genprog, monperrus2018automatic}. Traditional APR approaches mainly rely on heuristic search~\cite{le2011genprog, ghanbari2019prapr, qi2014strength, wen2018context, jiang2018shaping, yuan2018arja, martinez2016astor, saha2019harnessing, xin2017leveraging}, constraint solving~\cite{nguyen2013semfix, mechtaev2015directfix, mechtaev2016angelix, le2017s3, gao2021beyond, xuan2016nopol, xiong2017precise, afzal2019sosrepair}, or predefined templates~\cite{kim2013automatic, liu2019tbar, koyuncu2020fixminer, zhang2023gamma, hua2018towards, jiang2018shaping, saha2017elixir} to generate patches by mutating code, synthesizing logic, or applying fix patterns. Although these methods have laid the foundation for the field, they are often limited by restricted search spaces and a dependence on surface-level syntactic manipulations or manually crafted patterns~\cite{qi2015analysis, monperrus2018automatic}.

In recent years, the adoption of Large Language Models (LLMs) has transformed the landscape of APR, shifting the dominant paradigm from heuristic search to neural generation~\cite{xia2023automated, xia2022less}. Trained on massive-scale code corpora, LLMs demonstrate strong capabilities in code comprehension and logical reasoning~\cite{guo2024deepseek, roziere2023code, li2023starcoder, chen2021evaluating}. State-of-the-art approaches exploit these models through diverse strategies: ChatRepair~\cite{xia2024automated} utilizes conversational feedback for patch refinement; ThinkRepair~\cite{yin2024thinkrepair} and REINFIX~\cite{zhang2025repair} enhance repair performance via chain-of-thought reasoning and the retrieval of repair ingredients, respectively; while recent agentic frameworks, such as RepairAgent~\cite{bouzenia2025repairagent} and AdverIntent-Agent~\cite{ye2025adverintent}, automate the repair process by actively invoking external tools or utilizing adversarial feedback. These methods have shown substantial improvements over traditional techniques on standard benchmarks in both the number of fixed bugs and patch quality.

Despite recent advancements, current LLM-based approaches still face significant limitations. First, most existing methods rely primarily on static code analysis or coarse-grained dynamic feedback. While feedback-driven approaches, such as ChatRepair~\cite{xia2024automated} and Self-Debugging~\cite{chen2024teaching}, utilize standard error streams (stderr) or test outputs to guide the repair process, such information is often insufficient for complex bugs. These signals typically indicate where a failure occurred but fail to reveal why it happened, as they lack visibility into the intermediate execution values and data flow leading up to the error. Unlike human developers who inspect runtime states to trace the logic flaw, models operating without this detailed execution history often struggle to resolve silent failures. 
{Recent trace-based methods address this limitation by exposing intermediate values, branch decisions, and execution states to LLMs~\cite{haque2025towards, bouzenia2023tracefixer}. Debugger-assisted and step-by-step execution methods similarly inspect runtime states to diagnose and explain failures that are difficult to infer from source code alone~\cite{wang2026inspectcoder, zhong2024debug, kang2025explainable}. However, these methods generally use runtime information as additional prompt context or debugging feedback, rather than structuring the repair process around multiple trace-grounded hypotheses that are ranked, compared, and revised against shared runtime evidence.}

\begin{figure*}[ht]
\centering
\includegraphics[width=0.9\linewidth]{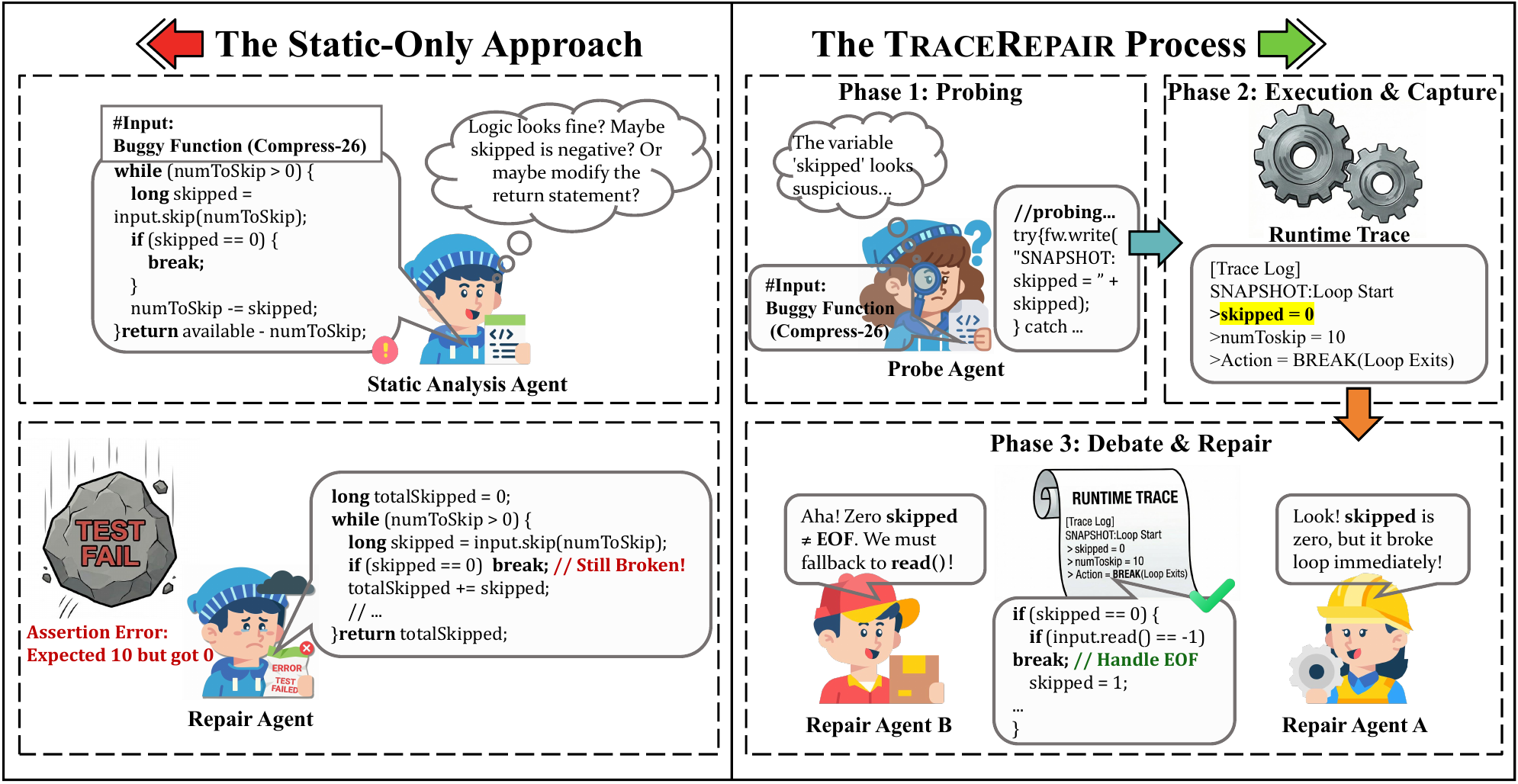}
\caption{Compress-26 repair comparison between the static-only approach (left) and \textsc{TraceRepair} (right). The static agent misattributes the fault to \texttt{numToSkip -= skipped}, a line never reached. \textsc{TraceRepair} captures \texttt{skipped=0, Action=BREAK} via the Probe Agent and guides Repair Agents to the correct \texttt{read()} fallback.}
\label{fig:motivating_example}
\vspace{-4mm}
\end{figure*}

Second, {runtime traces reduce diagnostic uncertainty but do not eliminate ambiguity among competing repair hypotheses. A single trace may still support multiple explanations, and several patches may appear plausible under the same observed behavior. Consequently,} relying on a single agent or simple feedback loops is often insufficient for robust repair. Research indicates that LLMs operating in isolation are prone to hallucination and struggle to self-correct erroneous reasoning~\cite{huang2024large, ji2023survey}. One common approach is large-scale sampling: generating and ranking multiple candidates to maximize the probability of finding a correct patch~\cite{xia2023automated, xia2022less}. However, this approach prioritizes quantity over quality, incurring high computational costs while increasing the risk of test-overfitting. While multi-agent debate has shown effectiveness in improving reasoning through cross-verification in other Natural Language Processing domains~\cite{du2023improving, chan2024chateval}, {its use for systematically comparing repair hypotheses against concrete runtime traces in APR remains less explored.}

Motivated by these observations, we propose \textsc{TraceRepair}, an agentic framework that combines runtime trace collection with multi-agent debate. {Unlike approaches that mainly use traces as additional prompt context or debugging feedback, \textsc{TraceRepair} uses runtime evidence to guide the construction and comparison of repair hypotheses.} Specifically, the framework employs an autonomous Probe Agent to inject logging statements into the buggy program, capturing detailed variable snapshots to {help identify} the root cause. Guided by these traces, \textsc{TraceRepair} initializes repair agents with defensive, causal, and semantic reasoning perspectives, enabling the generation of diverse fix candidates. {The generated candidates are first ranked using execution feedback, and the highest-ranked hypotheses then enter an iterative trace-guided debate. During the debate, the captured execution data is used to check whether candidate patches are consistent with the observed runtime states and test feedback; proposals contradicted by this evidence are challenged by peer agents.} Finally, a Judge Agent arbitrates the debate and synthesizes the {remaining evidence into a final patch}.

{Our evaluation focuses on the repair stage: given a buggy program, failure-triggering tests, and an executable environment, the task is to generate and validate patches.} We evaluated \textsc{TraceRepair} on the widely-used Defects4J benchmark (v1.2 and v2.0)~\cite{just2014defects4j} and a newly constructed  dataset \textsc{Recent-Java}. Experimental results show that, {under this evaluation setting,} \textsc{TraceRepair} achieves state-of-the-art repair effectiveness while substantially reducing token consumption compared with existing conversational and agentic baselines.

In summary, this paper makes the following contributions:

\textbf{Novel Framework:} We propose \textsc{TraceRepair}, a framework that injects diagnostic probes into buggy programs to collect runtime execution traces, then uses these traces {to guide candidate generation, execution-based ranking, and structured multi-agent debate}. Repair agents with distinct strategies generate and cross-validate patch candidates against the observed runtime state, {reducing the risk of accepting patches} that contradict actual program behavior.

\textbf{Extensive Evaluation:} We conduct an extensive evaluation on the Defects4J benchmark, where \textsc{TraceRepair} correctly fixes 392 bugs, outperforming existing LLM-based APR methods. To assess robustness against data leakage, we further construct a benchmark, \textsc{Recent-Java}, and show that \textsc{TraceRepair} maintains strong performance on unseen defects beyond model training data.

\textbf{Open Science:} We release the full implementation of \textsc{TraceRepair}, evaluation scripts, and all experimental data to support reproducibility and research. The replication package is anonymously available at:
\url{https://doi.org/10.5281/zenodo.19252356}.


\section{Motivation}
\label{sec:motivation}
A core limitation of current APR tools lies in their reliance on static analysis, which is often ineffective when bugs produce no explicit crash signals~\cite{monperrus2018automatic, qi2015analysis, 
wei2023copiloting}. We illustrate this through a real-world defect from Defects4J, showing where static diagnosis fails and how runtime instrumentation exposes the true fault.

\subsection{The Challenge of Silent Failures}
A fundamental challenge in APR is handling silent failures, which are logic errors where the program executes successfully but produces incorrect results~\cite{su2021fully}. In contrast to crashing bugs, where the stack trace restricts the search space to a specific execution path or line, silent failures often provide only a test assertion error (e.g., ``Expected X, but got Y'').

Without visibility into intermediate variable states or control flow decisions, the model must rely solely on the source code and the final failure message, often targeting code regions unrelated to the actual fault~\cite{ghanbari2022patch}.

\subsection{Motivating Example: Compress-26}
Figure~\ref{fig:motivating_example} illustrates a bug from the Commons-Compress project (Compress-26). The method aims to skip \texttt{numToSkip} bytes by calling \texttt{input.skip()}. The loop is designed to break if \texttt{skip()} returns 0.

\begin{figure*}[ht]
  \centering
  \includegraphics[width=0.9\textwidth]{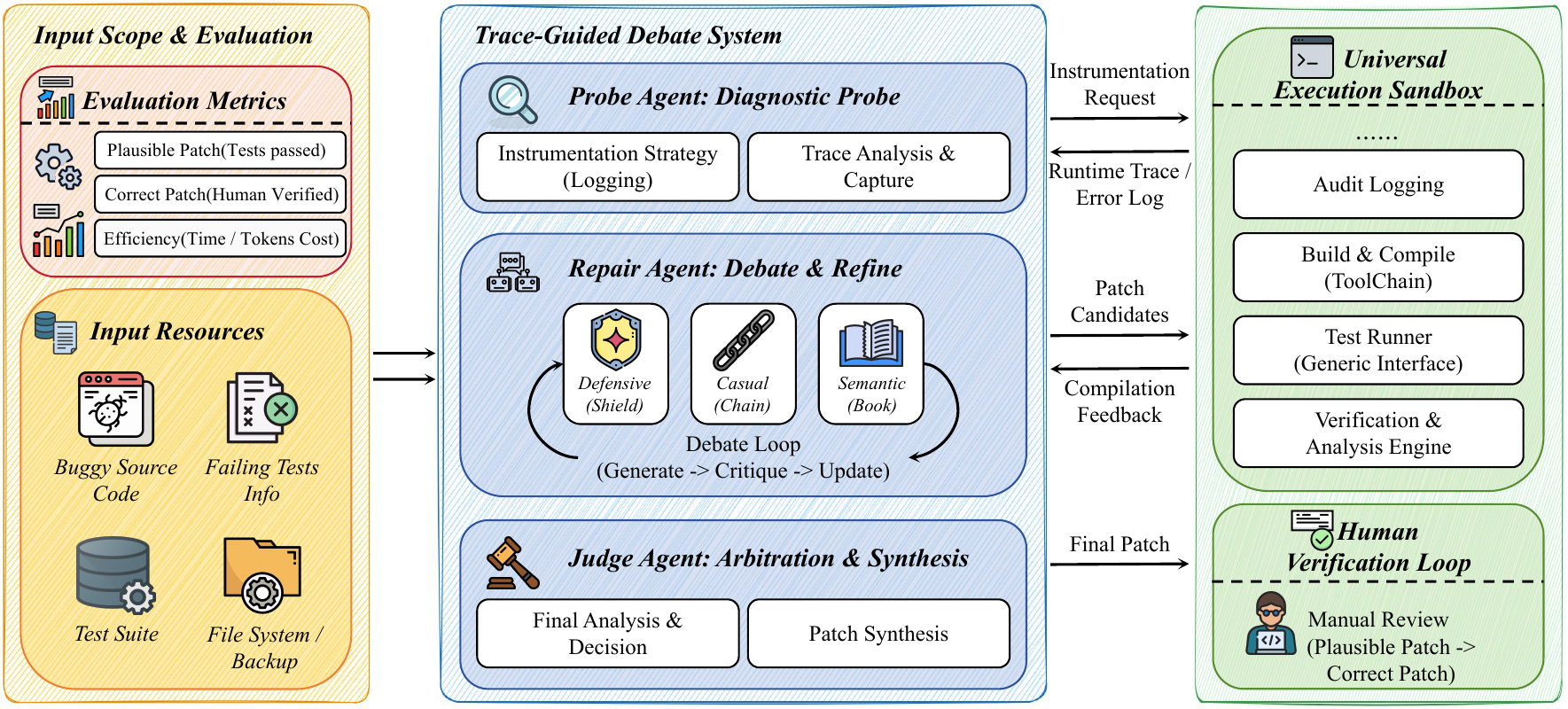}
  \caption{The Architecture of \textsc{TraceRepair}. The framework comprises a Trace-Guided Debate System interacting with a Universal Execution Sandbox.}
  \Description{A system architecture diagram showing the interaction between Input Scope, the Trace-Guided Debate System (Probe, Repair, Judge agents), and the Universal Execution Sandbox.}
  \label{fig:framework}
  \vspace{-4mm}
\end{figure*}

In the buggy version, when \texttt{input.skip()} returns 0, the loop breaks immediately. The regression test fails with an assertion error: ``Expected 10, but got 0''.

\textbf{Limitations of Static Diagnosis.}
Without runtime context, the static agent cannot determine which part of the loop actually executed. As shown in Figure~\ref{fig:motivating_example} (Left), it misattributes the fault to \texttt{numToSkip -= skipped} and proposes a patch targeting the subtraction logic.

This patch is invalid: the \texttt{break} statement fires before the arithmetic operation is ever reached. Without an execution trace, there is no way to verify this assumption.

\textbf{Resolution via \textsc{TraceRepair}.}
\textsc{TraceRepair} addresses this through dynamic diagnosis. The Probe Agent injects logging statements and captures the runtime trace \texttt{[skipped=0, ... Action=BR-} \texttt{EAK]}, as shown in Figure~\ref{fig:motivating_example} (Right).

The trace confirms the loop exited at the first iteration, ruling out any fault in the arithmetic logic. The return value of 0 instead points to a stalled stream—the correct fix is a fallback to \texttt{read()} for the remaining bytes.

\section{Methodology}
\label{sec:methodology}

We propose \textsc{TraceRepair}, a framework that augments static code reasoning with runtime execution evidence. As shown in Figure~\ref{fig:framework}, the system comprises a Trace-Guided Debate System operating over a Universal Execution Sandbox. The workflow proceeds in three phases:

\begin{enumerate}

\item \textbf{Phase 1: Diagnostic Probing.} The Probe Agent identifies the failure context and executes an instrumentation strategy. By inserting logging statements, the system collects test-specific runtime observations, including the values of selected variables and any branch or action markers recorded by the probes.

\item \textbf{Phase 2: Multi-Strategy Debate.} We initialize agents with distinct repair strategies: Defensive, Causal, and Semantic. These agents operate within a Generate, Critique and Update loop, {where the captured runtime observations and execution feedback are used to generate and review patch candidates}.

\item \textbf{Phase 3: Arbitration and Synthesis.} Finally, a Judge Agent analyzes {the summaries of the remaining proposals and the runtime observations} to resolve conflicts between strategies and synthesize a final patch, which is verified within the sandbox.
\end{enumerate}

\begin{algorithm}[htbp]
\caption{The \textsc{TraceRepair} Procedure}
\label{alg:tracerepair}
\small
\begin{algorithmic}[1]
\Require Buggy Program $P$, {Failure-triggering Tests $T_{\text{fail}}$, Candidate Budget $N$,} Max Rounds $R_{\text{max}}$
\Ensure Fixed Program $P_{\text{fix}}$ or $\bot$

\State \textbf{Phase 1: Diagnostic Probing}
\State $P_{\text{inst}}, \textit{status}
       \leftarrow \Call{Instrument}{P,{T_{\text{fail}}}}$
\If{$\textit{status}=\text{Success}$}
    \State $\tau\leftarrow
           \Call{ExecuteAndCapture}{P_{\text{inst}},{T_{\text{fail}}}}$
\Else
    \State $\tau\leftarrow\emptyset$
\EndIf

\State \textbf{Phase 2: Multi-Strategy Debate}
\State $\mathcal{H}\leftarrow\emptyset$
\State $\textit{Strategies}\leftarrow
       \{\text{Defensive},\text{Causal},\text{Semantic}\}$

\ForAll{$s\in\textit{Strategies}$}
    \State {$h_s\leftarrow\Call{BestOfN}{P,s,\tau,N}$}
    \If{{$\Call{Score}{h_s}=2$}}
        \State \Return {$h_s$}
    \EndIf
    \State $\mathcal{H}\leftarrow\mathcal{H}\cup\{h_s\}$
\EndFor

\For{$r\leftarrow1\dots R_{\text{max}}$}
    \State {$\mathcal{K}\leftarrow
           \Call{CollectCritiques}{\mathcal{H},\tau}$}
    \State {$\mathcal{H}\leftarrow
           \Call{ReviseAndEvaluate}{P,\mathcal{H},\mathcal{K},\tau}$}
    \If{{$\Call{TopScore}{\mathcal{H}}=2$}}
        \State \Return {$\Call{Top}{\mathcal{H}}$}
    \EndIf
\EndFor

\State \textbf{Phase 3: Arbitration and Synthesis}
\State {$M\leftarrow\Call{SummarizeCandidates}{\mathcal{H}}$}
\State {$P_{\text{final}}\leftarrow
       \Call{JudgeSynthesis}{P,M,\tau}$}
\If{{$P_{\text{final}}=\emptyset$}}
    \State {$P_{\text{final}}\leftarrow
           \Call{SelectByScoreAndRationale}{\mathcal{H}}$}
\EndIf
\If{$\Call{Validate}{P_{\text{final}}}=\text{PASS}$}
    \State \Return $P_{\text{final}}$
\Else
    \State \Return {$\bot$}
\EndIf

\end{algorithmic}
\end{algorithm}

\subsection{\textsc{TraceRepair} Algorithm Overview}
\label{sec:algorithm_summary}

Algorithm~\ref{alg:tracerepair} outlines the complete workflow of \textsc{TraceRepair}. 
\textit{1) Diagnostic Probing} (lines 2--7). The procedure begins by instrumenting the buggy program $P$ and executing it against {the failure-triggering tests $T_{\text{fail}}$} to collect {runtime observations} $\tau$ (lines 2--4). If instrumentation fails, $\tau$ is set to $\emptyset$ and the subsequent phases fall back to static reasoning (lines 5--6). 
\textit{2) Multi-Strategy Debate} {(lines 8--24)}. Repair agents are initialized, each adopting a distinct reasoning strategy: Defensive, Causal, and Semantic. Each agent independently {generates and evaluates $N$ candidate patches through \textsc{BestOfN}. If a candidate obtains score 2, it is returned immediately; otherwise, the selected candidate is used as the initial hypothesis for that strategy}, forming the hypothesis set $\mathcal{H}$ {(lines 9--17)}. The agents then enter an iterative debate for at most $R_{\max}$ rounds {(lines 18--24)}. {In each round, the agents critique the current hypotheses based on the runtime observations and execution feedback. Each hypothesis is then revised and evaluated again (lines 19--20). If an updated hypothesis obtains score 2, the highest-scoring hypothesis is returned immediately (lines 21--23).} 
\textit{3) Arbitration and Synthesis} {(lines 25--35)}. After the debate concludes, {the Judge Agent synthesizes a final candidate $P_{\text{final}}$ from the summaries of the remaining hypotheses and the runtime observations $\tau$ (lines 26--27). If the Judge does not produce usable code, the remaining hypothesis with the highest execution score is used as a fallback, with ties broken by rationale length (lines 28--30). The resulting candidate is then validated. If it passes validation, it is returned as the final patch; otherwise, the procedure returns $\bot$ (lines 31--35).}

\vspace{-2mm}

\subsection{Phase 1: Diagnostic Probing}
\label{sec:phase1}

Static repair often fails due to a lack of visibility into the execution state. To overcome this, the Probe Agent uses diagnostic instrumentation to expose internal variable states.

\textbf{Targeted Variable Identification.}
Rather than randomly probing the codebase, the agent analyzes {the benchmark-provided target method, failure-triggering test names, and error messages} to identify critical variables ($V_{\text{crit}}$) {within the target method}. These variables, such as loop counters, conditional flags, or buffer indices, are selected based on their potential to reveal the error state. For instance, in the Compress-26 case, the agent identifies the return value of \texttt{input.skip()} as the key variable determining control flow.

{The Probe Agent does not generate runtime facts or repair the program in this phase. It only selects variables and generates an instrumented method. The runtime observations are produced by building and executing the instrumented program against the failure-triggering tests in the Universal Execution Sandbox.}

\textbf{Safe Instrumentation Strategy.}
{In our implementation, the Probe Agent acts as a recorder rather than a fixer. It only inserts logging statements around suspicious locations while keeping the original program logic unchanged. The logging code follows a fixed \texttt{try-catch} template to record selected variable values and handle possible logging exceptions.}

{After generating the instrumented method, \textsc{TraceRepair} compiles the instrumented project in the Universal Execution Sandbox before collecting traces. If the compilation fails, the modification is rolled back and the Probe Agent generates a new instrumentation. This process ensures that only valid instrumentation is used for trace collection.}


\textbf{Trace Volume Management.}
Runtime execution traces, particularly those emerging from loops or deep recursion, can rapidly exceed the context window limits of LLMs. To mitigate this without losing diagnostic fidelity, \textsc{TraceRepair} implements a \textit{Tail-Biased Truncation} strategy.

{We apply two controls to limit the trace size.} First, at the execution level, the instrumented probe includes a guard clause that {stops writing new entries} if the file size exceeds 10MB; this prevents storage exhaustion during infinite loops. Second, during prompt construction, we prioritize the immediate pre-failure state. If a trace log exceeds {2,000 bytes}, the system discards the initial entries and retains only the {last 2,000 bytes}. This follows from the observation that for crash-type and assertion-failure bugs, the most critical causal evidence typically resides in the final state transitions immediately preceding the termination.

\textbf{Trace Extraction.}
{The Universal Execution Sandbox executes the instrumented program ($P_{\text{inst}}$) against each failure-triggering test to collect the runtime observations ($\tau$).} {The collected evidence is represented as an ordered set of test-specific diagnostic logs:}
\begin{equation*}
{\tau=\langle (t,\ell_t)\mid t\in T_{\text{fail}}\rangle}
\end{equation*}
{where $\ell_t$ denotes the retained diagnostic log produced by test $t$. Depending on the generated probes, a log may contain selected variable values, condition values, or execution markers.}

{The goal of trace collection in \textsc{TraceRepair} is to capture runtime evidence directly related to the observed failure. Based on failure-triggering tests, the Probe Agent identifies suspicious locations and collects local runtime observations around these locations, including critical variable values and branch or action information recorded by the probes. These traces reveal the program states and execution behaviors related to the failure, allowing repair agents to evaluate whether repair hypotheses are consistent with the observed program behavior.}

\subsection{Phase 2: Multi-Strategy Debate}
\label{sec:phase2}

To avoid convergence on incorrect but plausible solutions, \textsc{TraceRepair} does not rely on a monolithic generator. Instead, it orchestrates a committee of agents that engage in a structured debate. This phase proceeds in two stages: strategy initialization and the trace-guided debate loop.

\textbf{Strategy Initialization.}
We initialize a diverse hypothesis space ($\mathcal{H}$) by employing agents equipped with distinct reasoning strategies. Conditioned on the buggy code ($P_{\text{bug}}$), {failure information,} and the runtime trace ($\tau$), each agent generates candidate patches from a distinct perspective:

\begin{itemize}
\item \textbf{Defensive Strategy.} Prioritizes system robustness. Guided by anomalies in the trace, it {checks whether null checks, boundary checks, or fallback logic are needed} to prevent state corruption.
\item \textbf{Causal Strategy.} Focuses on error propagation. It analyzes {the recorded runtime values} to identify {why an incorrect state is produced} and repairs the underlying control flow {or computation} logic.
\item \textbf{Semantic Strategy.} Ensures specification adherence. It scrutinizes the code against API contracts {, comments, failure information,} and domain-specific constraints.
\end{itemize}

To seed the debate with high-quality candidates, we employ a Best of $N$ Selection mechanism. Candidates undergo pre-evaluation in the sandbox. {A candidate receives a score of 0 if the project fails to build, a score of 1 if it builds but fails either the designated failure-triggering test or the subsequent validation tests, and a score of 2 if it passes both stages. A candidate with a score of 2 is accepted immediately. Otherwise, the highest-scoring candidate from each strategy is selected. Ties are broken by rationale length and then by generation order.}

\textbf{The Trace-Guided Debate Loop.}
In traditional multi-agent frameworks, agents often reach a false consensus on incorrect patches. \textsc{TraceRepair} {reduces this risk by asking agents to review failed repair hypotheses using both runtime observations and execution feedback}. The debate iterates between Proposers and Critics.

For every candidate patch $h_{\text{def}}$ (Defender), an opposing agent $h_{\text{atk}}$ (Attacker) {reviews the proposal using the runtime observations $\tau$ and its execution feedback $e_{\text{def}}$}. We formally define the critique generation function $\mathcal{F}_{\text{crit}}$ as:
\begin{equation}
{k=\mathcal{F}_{\text{crit}}(h_{\text{atk}},h_{\text{def}},\tau,e_{\text{def}})}
\end{equation}
{The critique is produced by an LLM-based reviewer rather than a formal consistency checker. The reviewer receives the buggy code, the runtime observations, the proposal's repair logic and rationale, and its execution feedback. It explains why the proposal may have failed and gives concrete suggestions for the next attempt.}

The Defender responds to the critique set $\mathcal{K}$ by synthesizing a refined patch. This update step {generates a new complete method}, defined as:
\begin{equation}
{h_{\text{new}}=\mathcal{F}_{\text{update}}(P_{\text{bug}},\mathcal{K}[h_{\text{def}}],\tau)}
\end{equation}

{A failed candidate is transformed into a new repair attempt in two steps. First, the other strategy agents review its repair logic and execution feedback. The feedback may describe a build failure, a failure-triggering test failure, or a failure in the subsequent validation tests. Second, the update agent receives the original buggy method, the runtime observations, and the peer critiques, and rewrites the complete target method under the same repair strategy. The current candidate and its execution result affect the new proposal through these critiques.}

{The rewritten method is built and tested again in the Universal Execution Sandbox. If it receives score 2, it is accepted immediately. Otherwise, it replaces the previous proposal and enters the next debate round. The replacement does not guarantee that the execution score increases in every round. It is a new repair attempt informed by the failure of the previous proposal.} This iterative process continues until {a candidate passes the implemented validation procedure} or the maximum {round} limit is exceeded.

\subsection{Phase 3: Arbitration and Synthesis}
\label{sec:phase3}

This phase is invoked when the debate loop reaches the maximum number of rounds ($R_{\text{max}}$) without producing a validated patch. The Judge Agent then performs a final synthesis procedure.

\textbf{{Summary-Based Synthesis.}}
{The Judge Agent receives the original buggy method, the runtime observations $\tau$, and a compact summary of each remaining hypothesis, including its strategy, execution score, repair logic, and rationale. Based on these inputs, the Judge generates a new complete method rather than directly selecting or merging the candidate implementations.}

{The generated method is inserted into the target program and evaluated in the Universal Execution Sandbox. If it passes the implemented validation procedure, it is returned as the final patch. Otherwise, \textsc{TraceRepair} falls back to the remaining hypothesis with the highest execution score, where ties are resolved based on the rationale length. The selected candidate is then validated before being returned. If the validation fails, the procedure reports a repair failure.}

\section{Experiment Setup}
\label{sec:setup}

\subsection{Research Questions}
\label{sec:rqs}

To evaluate \textsc{TraceRepair}, we design our experiments around four research questions:

\begin{itemize}
    \item \textbf{RQ1: Can the framework outperform existing baselines in generating correct patches?}
    
    We evaluate on Defects4J (v1.2 and v2.0) under both Perfect Fault Localization (PFL) and Method-Level Fault Localization (MFL) settings, comparing against the current best-performing baselines to comprehensively assess the framework's repair capability.
    
    \item \textbf{RQ2: How much does each component contribute to the overall repair performance?} 
    
    We conduct an ablation study by removing runtime traces and multi-agent debate separately, measuring the individual contribution of each component to the final repair results.
    
    \item \textbf{RQ3: Can the framework generalize to defects beyond its training distribution?}
    
    We evaluate on a newly constructed dataset comprising bugs collected from repositories active after the training cutoff of the underlying models, assessing whether the framework's repair capability stems from genuine dynamic reasoning rather than memorization of previously seen fixes.
    
    \item \textbf{RQ4: Does \textsc{TraceRepair} achieve its repair gains without excessive computational overhead?}
    
    We measure token consumption and financial cost per fix, comparing \textsc{TraceRepair} against conversational and agentic baselines to evaluate the cost-effectiveness of the framework.
\end{itemize}

\subsection{Benchmarks}
\label{sec:benchmarks}

To ensure a rigorous evaluation, we conduct experiments on three benchmarks: Defects4J v1.2, Defects4J v2.0 and \textsc{Recent-Java}.

\textbf{Defects4J benchmark.} We utilize both v1.2 and v2.0 of the Defects4J benchmark, which contain real-world faults from large-scale Java projects (e.g., Commons-Lang, JFreeChart). This benchmark is widely adopted by existing LLM-based APR approaches, ensuring a fair comparison with state-of-the-art baselines. As shown in Table~\ref{tab:dataset_stats}, we categorize the defects based on repair complexity: Single-Function (SF) bugs, which require modifications within a single method, and Multi-Function (MF) bugs, which involve synchronized changes across multiple methods. We further detail Single-Hunk (SH) and Single-Line (SL) bugs to assess performance on localized faults. {Following the evaluation scope commonly adopted in prior LLM-based APR studies~\cite{xia2024automated,zhang2025repair}, we evaluate 673 single-file Defects4J bugs with reproducible failure-triggering tests, excluding cases outside the supported repair scope or not reproducible in our environment.}

\begin{table}[h]
  \caption{Statistics of Defects4J.}
  \label{tab:dataset_stats}
  \footnotesize
  \centering
  \begin{tabular*}{\columnwidth}{@{\extracolsep{\fill}} l c c c c @{}}
    \toprule
    \textbf{Benchmarks} & \textbf{\#MF Bugs} & \textbf{\#SF Bugs} & \textbf{\#SH Bugs} & \textbf{\#SL Bugs} \\
    \midrule
    Defects4J v1.2   & 136 & 255 & 154 & 80 \\
    Defects4J v2.0   & 210 & 228 & 159 & 78 \\
    \bottomrule
  \end{tabular*}
  \vspace{-4mm}
\end{table}

\textbf{\textsc{Recent-Java}.}
Data leakage poses a critical threat to the validity of LLM-based repair evaluation, while constructing post-cutoff benchmarks is inherently challenging due to the scarcity of recent, high-quality defects. {To reduce this risk, we construct \textsc{Recent-Java}, a post-cutoff benchmark for evaluating different backbone models. We select GPT-3.5 and DeepSeek-Coder as representative models, whose reported training cutoffs are September 2021 and March 2023, respectively. To place all defects after the later reported cutoff, we collect bug-fixing commits dated between September 2023 and December 2025 from three Java projects: Jsoup, Commons-Lang, and Commons-Compress.}

{For each candidate commit, we retain only bugs for which the buggy and fixed methods can be identified, the developer patch is confined to a single method, the triggering failure is reproducible, and the Maven backport validation succeeds. This process results in 21 bugs. To assess the quality of \textsc{Recent-Java}, we compare it with the Defects4J single-function subset used in our experiments in terms of buggy-method size, patch size, cyclomatic complexity, and validation test suite size. As shown in Table~\ref{tab:recent_java_quality}, \textsc{Recent-Java} has an average buggy method size of 22.38 LOC, an average patch size of 4.81 edited lines, an average cyclomatic complexity of 7.71, and an average validation test suite size of 11,310 tests. The corresponding values for the Defects4J subset are 41.84 LOC, 5.29 edited lines, 11.29, and 2,614 tests, respectively. Thus, \textsc{Recent-Java} contains shorter and less complex buggy methods while requiring a comparable amount of code modification. Its larger validation test suites result from the use of recent snapshots of actively maintained projects, whose regression tests have continuously accumulated as the projects evolved. In other words, despite their smaller code contexts, these bugs require a comparable amount of code modification, and the generated patches are validated against substantially larger regression test suites. Therefore, \textsc{Recent-Java} provides suitable evaluation cases for APR methods.}

\begin{table}[h]
\vspace{-2mm}
\centering
\caption{{Comparison of \textsc{Recent-Java} with the Defects4J SF bugs.}}
\label{tab:recent_java_quality}
\footnotesize
\setlength{\tabcolsep}{3pt}
\begin{tabular*}{\columnwidth}{
    @{\extracolsep{\fill}} llcc @{}
}
\toprule
{\textbf{Category}}
& {\textbf{Metric}}
& \shortstack{{\textbf{Defects4J SF}}}
& {\textbf{\textsc{Recent-Java}}} \\
\midrule

\multirow{2}{*}{{Code size}}
& {Avg. buggy method LOC}
& {41.84}
& {22.38} \\

& {Avg. patch line edits}
& {5.29}
& {4.81} \\
\midrule

{Complexity}
& {Avg. cyclomatic complexity}
& {11.29}
& {7.71} \\
\midrule

{Number of tests}
& {Avg. number of validation tests}
& {2614}
& {11310} \\
\bottomrule
\end{tabular*}
\vspace{-3mm}

\end{table}

\subsection{Baselines}
\label{sec:baselines}

To evaluate the effectiveness of \textsc{TraceRepair}, we compare it against ten state-of-the-art LLM-based APR approaches, covering five representative categories: (1) feedback-driven methods, including ChatRepair~\cite{xia2024automated} and D4C~\cite{xu2025aligning}; (2) retrieval-augmented methods, including REINFIX~\cite{zhang2025repair}; (3) agentic and reasoning-based methods, including RepairAgent~\cite{bouzenia2025repairagent}, ThinkRepair~\cite{yin2024thinkrepair}, and AdverIntentAgent~\cite{ye2025adverintent}; (4) template-based and hybrid methods, including GiantRepair~\cite{li2025hybrid} and FitRepair~\cite{xia2023plastic}; and (5) direct generative methods, including GAMMA~\cite{zhang2023gamma} and Codex~\cite{chen2021evaluating}.


\subsection{Evaluation Metrics}
\label{sec:metrics}

We evaluate performance using two standard metrics: \textit{plausible patches}, which satisfy the full test suite (passing both $T_{\text{fail}}$ and $T_{\text{reg}}$), and \textit{correct patches}, which are semantically equivalent to the developer's patches. Mere test passage does not guarantee correctness due to the overfitting phenomenon, where patches exploit weak test specifications without fixing the underlying logic~\cite{smith2015cure}. To rigorously mitigate this threat, we adopted the manual verification protocol established in prior studies~\cite{jiang2021cure, zhu2021syntax, ye2022neural}. Specifically, two authors {with over five years of Java programming experience} independently inspected every plausible patch {by examining the generated patch, the developer patch, the buggy method and its source location, the triggering test cases and failure messages, and the corresponding issue description when available}. {A plausible patch was considered correct if it was either syntactically identical to the developer patch, except for non-semantic differences such as formatting, or semantically equivalent to the developer patch, implementing the intended fix without introducing evident behavioral changes or regressions. When the two authors initially disagreed, they re-examined and discussed the patch based on the available evidence. If disagreement remained after discussion, the patch was not counted as correct.}

\subsection{Implementation Details}
\label{sec:implementation}

\textsc{TraceRepair} is implemented in Python and interacts with LLMs via the standard OpenAI API interface. The framework is model-agnostic and can be seamlessly integrated with any LLM backend. To evaluate its generalizability across different model families, we instantiate it with two representative backbones: DeepSeek-V3.2 and GPT-3.5-Turbo, hereafter referred to as \textsc{TraceRepair}$_{DSV3.2}$ and \textsc{TraceRepair}$_{GPT3.5}$, respectively. To promote search space diversity, we configured the generation agents with a temperature of 1.0. We employed a sampling strategy with a maximum budget of 40 LLM invocations per bug, incorporating an early-stopping mechanism that terminates generation once a plausible patch is identified. We enforced a strict 300-second timeout for all test executions to prevent non-terminating executions caused by faulty patches. The multi-agent debate is limited to 3 rounds.

\section{Evaluation Results}
\label{sec:results}

\subsection{RQ1: Effectiveness}
\label{sec:rq1}


\begin{table*}[h]
\caption{Repair performance (Correct / Plausible) on Defects4J under Perfect Fault Localization.}
\label{tab:main_results}
\footnotesize
\centering
\renewcommand{\arraystretch}{1.05}
\resizebox{\textwidth}{!}{%
\begin{tabular}{c | c c | c c c c c c c c c} 
\toprule
\textbf{Benchmark} & \textbf{\textsc{TraceRepair}$_{DSV3.2}$} & \textbf{\textsc{TraceRepair}$_{GPT3.5}$} & \textbf{REINFIX$_{GPT4o}$} & \textbf{REINFIX$_{GPT3.5}$} & \textbf{GiantRepair} & \textbf{ChatRepair} & \textbf{RepairAgent} & \textbf{ThinkRepair} & \textbf{AdverIntent-Agent} & \textbf{FitRepair} & \textbf{GAMMA} \\
\textit{Backbone} & \textit{DeepSeek-V3.2} & \textit{GPT-3.5} & \textit{GPT-4o} & \textit{GPT-3.5} & \textit{GPT-4} & \textit{GPT-3.5} & \textit{GPT-3.5} & \textit{GPT-3.5} & \textit{GPT-4} & \textit{Code-T5} & \textit{UniXcoder} \\
\textit{Sample Times} & 40 & 40 & 45 & 45 & 500 & 500 & 117 & 125 & - & 4000 & 250 \\
\midrule
Chart        & 19 / 19 & 15 / 16 & 18 / 20 & 16 / 17 & 8 / - & 15 / - & 11 / 14 & 11 / - & - & 8 / - & 11 / 11 \\
Closure      & 70 / 77 & 30 / 36 & 40 / 50 & 30 / 37 & 32 / - & 37 / - & 25 / 25 & 31 / - & - & 29 / - & 24 / 26 \\
Lang         & 37 / 44 & 19 / 26 & 33 / 47 & 26 / 33 & 14 / - & 21 / - & 17 / 17 & 19 / - & - & 19 / - & 16 / 25 \\
Math         & 66 / 78 & 35 / 43 & 39 / 68 & 35 / 52 & 26 / - & 32 / - & 29 / 29 & 27 / - & - & 24 / - & 25 / 31 \\
Mockito      & 19 / 21 & 6 / 7   & 10 / 11 & 8 / 9   & 6 / -  & 6 / -  & 6 / 6   & 6 / -  & - & 6 / - & 3 / 3 \\
Time         & 6 / 9   & 5 / 6   & 6 / 11  & 3 / 4   & 1 / -  & 3 / -  & 2 / 3   & 4 / -  & - & 3 / - & 3 / 5 \\
\midrule
D4J v1.2 Total & 217 / 248 & 110 / 134 & 146 / 207 & 118 / 152 & 87 / - & 114 / - & 90 / 94 & 98 / - & - & 89 / - & 82 / 101 \\
\midrule
D4J v2.0 Total & 175 / 204 & 114 / 130 & 145 / 190 & 123 / 147 & 84 / - & 48 / - & 74 / 92 & 107 / - & - & 44 / - & 45 / - \\
\midrule
\textbf{Total} & \textbf{392 / 452} & \textbf{224 / 264} & 291 / 397 & 241 / 299 & 171 / - & 162 / - & 164 / 186 & 205 / - & 141 / 180 & 133 / - & 127 / - \\
\bottomrule
\end{tabular}%
}
\vspace{-2mm}
\end{table*}

We first evaluate \textsc{TraceRepair} on Defects4J (v1.2 and v2.0) under the PFL setting, where the exact buggy line is provided as input. Following established practice in LLM-based APR research~\cite{xia2024automated, yin2024thinkrepair, zhang2025repair, bouzenia2025repairagent}, this setting isolates patch generation from localization errors, enabling fair and direct comparison across methods. As shown in Table~\ref{tab:main_results}, \textsc{TraceRepair}$_{DSV3.2}$ correctly resolves \textbf{392} defects (217 on v1.2 and 175 on v2.0), outperforming all compared methods in terms of the number of correctly fixed defects.

Notably, \textsc{TraceRepair}$_{DSV3.2}$ is built upon the DeepSeek-V3.2 model, and such a comparison may partially reflect differences in backbone model capacity rather than the repair framework itself. To control for this factor, we further evaluate our framework using GPT-3.5 as the backbone, which is widely adopted across prior APR approaches and provides a common basis for comparison. Under this controlled setting, \textsc{TraceRepair}$_{GPT3.5}$ achieves 224 correct fixes, outperforming most baselines, including several methods built on stronger models such as GPT-4. This result suggests that the observed improvements primarily stem from the proposed framework rather than the choice of backbone model. The only exception is REINFIX, which we analyze in detail below.


\textbf{Comparison with Conversational and Agentic Methods.} \textsc{TraceRepair}$_{GPT3.5}$ fixes 224 bugs, compared to 162 for ChatRepair and 164 for RepairAgent under the same backbone. Both baselines rely on test feedback or tool-assisted analysis to guide repair, which primarily indicates where a failure occurs but provides limited insight into the program’s runtime state. In contrast, execution traces explicitly capture fine-grained program state transitions, offering richer semantic signals that enable more accurate diagnosis and repair.

\textbf{Comparison with REINFIX.} Under the same GPT-3.5-turbo backbone, \textsc{TraceRepair}$_{GPT3.5}$ correctly fixes 224 bugs, compared to 241 for REINFIX$_{GPT3.5}$. While REINFIX$_{GPT3.5}$ achieves slightly higher performance, the gap is relatively modest. Notably, REINFIX explicitly incorporates retrieval-augmented generation to provide relevant fix patterns at inference time, offering strong priors that can directly guide patch construction. In contrast, \textsc{TraceRepair} does not rely on external retrieval but instead leverages execution traces to model fine-grained program state transitions. This design emphasizes dynamic reasoning over pattern reuse, which may be less advantageous in cases where similar fixes can be readily retrieved, but provides complementary strengths in scenarios requiring deeper semantic understanding.

\begin{figure*}[h]
    \centering
    \includegraphics[width=0.95\textwidth]{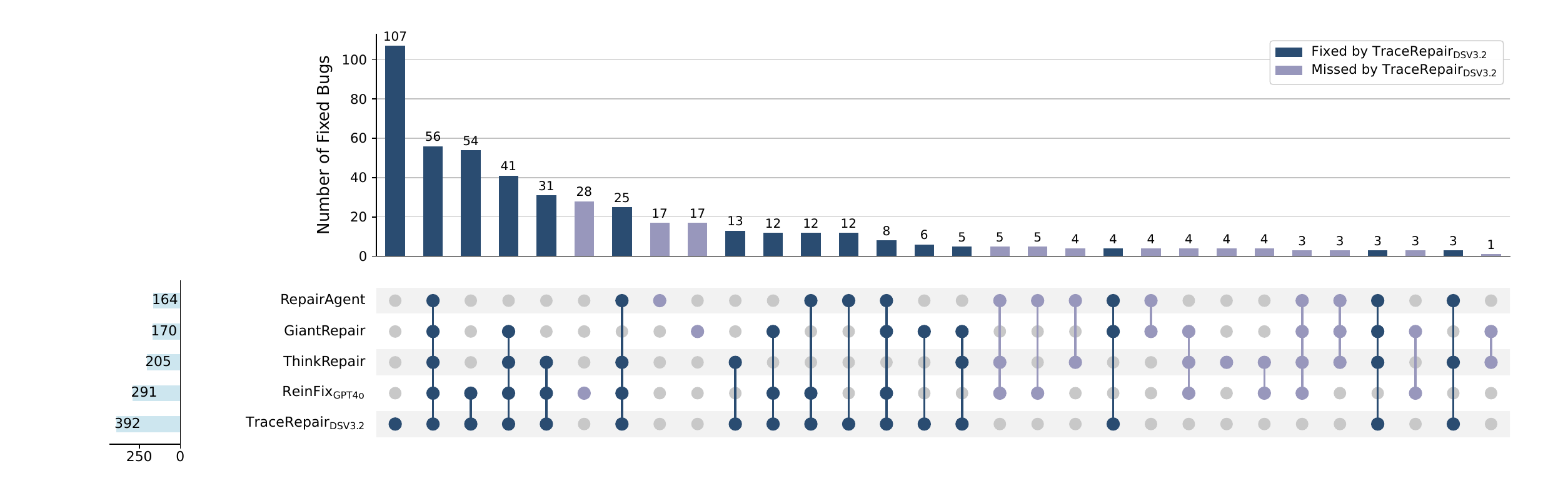} 
    \caption{Overlap analysis of correct fixes between \textsc{TraceRepair}$_{DSV3.2}$ and four representative baselines. Vertical bars show the number of bugs in each intersection set, and the leftmost bar represents bugs fixed exclusively by a single method.}
    \label{fig:overlap_analysis}
    \vspace{-3mm}
\end{figure*}

\textbf{Uniqueness and Difficulty Analysis.}
Raw fix counts alone may obscure performance differences on more challenging defects. To provide a finer-grained analysis, we examine the overlap of correctly fixed bugs between \textsc{TraceRepair}$_{DSV3.2}$ and four representative baselines, as shown in Figure~\ref{fig:overlap_analysis}. A total of 56 defects are resolved by all methods, indicating a shared subset of relatively straightforward bugs. More importantly, \textsc{TraceRepair}$_{DSV3.2}$ uniquely fixes 107 defects that are missed by all other baselines, substantially exceeding the unique contribution of REINFIX$_{GPT4o}$, which fixes 28 such defects. These uniquely resolved cases predominantly involve logic-dependent failures, where static signals are often insufficient to characterize the underlying issue. In contrast, execution traces provide explicit evidence of program state transitions, enabling more precise diagnosis and repair in such scenarios.


Table~\ref{tab:complexity_comparison} presents repair performance across two bug types: single-function (SF) bugs confined to a single method, and multi-function (MF) bugs requiring coordinated changes across multiple methods. \textsc{TraceRepair}$_{GPT3.5}$ achieves consistently strong performance in both settings, fixing 94 SF and 16 MF bugs on v1.2, and 101 SF and 13 MF bugs on v2.0. On MF bugs, which are substantially more challenging, \textsc{TraceRepair}$_{GPT3.5}$ fixes 29 defects in total, outperforming RepairAgent with 13 fixes and slightly surpassing REINFIX$_{GPT3.5}$ with 28 fixes. These results highlight the advantage of our approach in handling cross-method dependencies. MF bugs typically involve error propagation across multiple functions, where static analysis often fails to capture the full execution context. In contrast, execution traces explicitly model inter-procedural data flow and state transitions, enabling the model to produce globally consistent repairs across affected methods.

\begin{table}[htbp]
  \caption{Distribution of correct fixes across SF and MF bugs.}
  \label{tab:complexity_comparison}
  \footnotesize
  \centering
  \renewcommand{\arraystretch}{1.0}
  \begin{tabular*}{\columnwidth}{@{\extracolsep{\fill}} l c c c c @{}}
    \toprule
    \multirow{2}{*}{\textbf{Approach}} & \multicolumn{2}{c}{\textbf{Defects4J v1.2}} & \multicolumn{2}{c}{\textbf{Defects4J v2.0}} \\
    \cmidrule(lr){2-3} \cmidrule(lr){4-5}
    & \textbf{MF} & \textbf{SF} & \textbf{MF} & \textbf{SF} \\
    \midrule
    ChatRepair & - & 76 & - & - \\
    RepairAgent & 7 & 83 & 6 & 68 \\
    REINFIX$_{GPT3.5}$ & 14 & 104 & 14 & 109 \\
    \midrule
    \textbf{\textsc{TraceRepair}$_{GPT3.5}$} & 16 & 94 & 13 & 101 \\
    \textbf{\textsc{TraceRepair}$_{DSV3.2}$} & \textbf{40} & \textbf{177} & \textbf{19} & \textbf{156} \\
    \bottomrule
  \end{tabular*}
  \vspace{-2mm}
\end{table}

In addition to the PFL setting, we further evaluate our approach under Method-Level Fault Localization to better reflect real-world deployment conditions, where precise bug locations are typically unavailable and localization tools only provide coarse-grained scopes. We compare against baselines that have reported results under Method-Level Fault Localization on Defects4J v1.2 SF bugs. As shown in Table~\ref{tab:localization_robustness}, static approaches degrade substantially when the localization scope is expanded. Codex drops by 36.36\% and ThinkRepair by 18.37\%, as both rely on a narrow input scope to effectively constrain patch generation. In contrast, \textsc{TraceRepair}$_{GPT3.5}$ limits the drop to 10.64\%, while \textsc{TraceRepair}$_{DSV3.2}$ exhibits minimal degradation of only 3.39\%, decreasing from 177 to 171 correct fixes. This robustness stems from the use of execution traces, which capture fine-grained program state changes within the method body and provide an implicit localization signal. Even when the externally provided scope spans the entire method, these traces enable the model to identify the relevant faulty regions and generate accurate repairs.


\begin{table}[htbp]
  \caption{Performance drop when fault localization is degraded from PFL to MFL (Defects4J v1.2 SF bugs).}
  \label{tab:localization_robustness}
  \footnotesize
  \centering
  \begin{tabular*}{\columnwidth}{@{\extracolsep{\fill}} l c c c @{}}
    \toprule
    \textbf{Approach} & \textbf{PFL} & \textbf{MFL} & \textbf{Drop Rate} \\
    \midrule
    Codex & 99 & 63 & 36.36\% \\
    ThinkRepair & 98 & 80 & 18.37\% \\
    D4C & - & 84 & - \\
    GiantRepair & 87 & 64 & 26.4\% \\
    \midrule
    \textbf{\textsc{TraceRepair}$_{GPT3.5}$} & 94 & 84 & \textbf{10.64\%} \\
    \textbf{\textsc{TraceRepair}$_{DSV3.2}$} & \textbf{177} & \textbf{171} & \textbf{3.39\%} \\
    \bottomrule
  \end{tabular*}
\end{table}

\noindent\fbox{
  \parbox{0.95\linewidth}{
    \textbf{Summary of RQ1:} \textsc{TraceRepair} achieves  state-of-the-art results, especially on complex MF bugs. Using runtime traces for localization, it avoids the severe performance drops seen in static tools under MFL. It also performs closely to retrieval-augmented methods without needing external databases.
  }
}

\subsection{RQ2: Component Analysis}
\label{sec:rq2}

This section breaks down how each component of \textsc{TraceRepair} contributes to repair performance. Due to economic constraints, ablation experiments are conducted primarily on \textsc{TraceRepair}$_{DSV3.2}$ using Defects4J v1.2 (SF bugs). We examine two aspects: the cumulative gains from iterative debate, and the individual utility of runtime traces versus multi-agent collaboration.

\begin{figure}[h]
  \centering
  \includegraphics[width=0.48\textwidth]{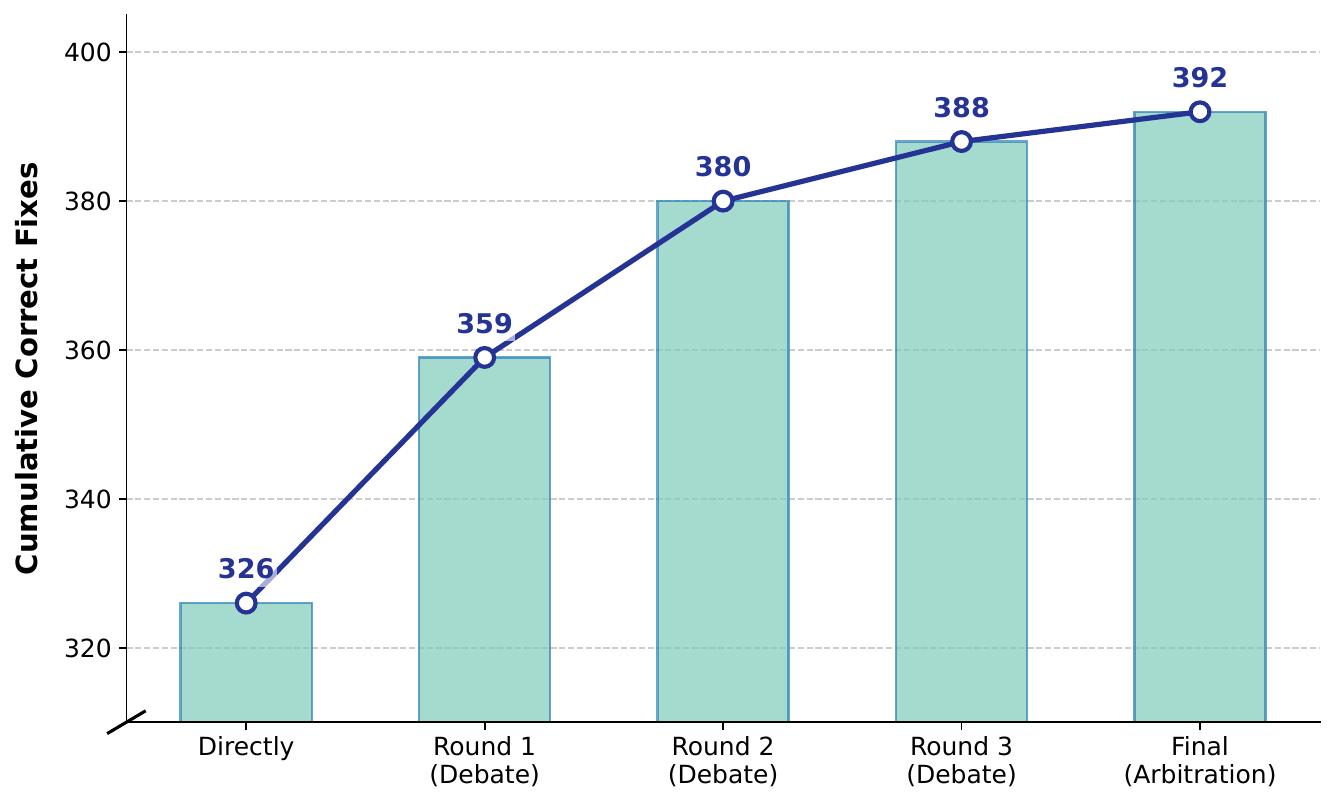} 
  \caption{Cumulative correct fixes across stages in \textsc{TraceRepair}$_{DSV3.2}$.}
  \label{fig:rq2_trend}
  \vspace{-4mm}
\end{figure}

\textbf{Cumulative Impact of Debate.}
Figure~\ref{fig:rq2_trend} illustrates the cumulative number of correct fixes across stages in \textsc{TraceRepair}$_{DSV3.2}$. Direct generation produces 326 correct fixes. Subsequent rounds progressively improve performance, with Round 1 contributing the largest gain of 33 fixes, followed by Round 2 with 21 fixes, Round 3 with 8 fixes, and a final arbitration stage adding 4 fixes, reaching a total of 392. Overall, the debate mechanism recovers 66 additional fixes over one-shot generation, corresponding to a 20.2 percent improvement. The diminishing gains across later rounds suggest that earlier iterations capture the majority of easily correctable errors, while subsequent rounds focus on refining more challenging cases.


\textbf{Ablation Study.}
To assess the contribution of each component, we conduct an ablation study that isolates runtime traces and the debate mechanism, and further removes each debate strategy individually. This setup evaluates whether the two core components are independently effective and whether all three strategies are necessary within the debate workflow. Experiments are conducted on Defects4J v1.2 SF bugs using \textsc{TraceRepair}$_{DSV3.2}$, with results summarized in Table~\ref{tab:ablation_study}. Removing runtime traces (\textsc{TraceRepair}$_{NT}$) reduces performance to 154 correct fixes, yet still yields a 50.9\% improvement over the baseline, indicating that the debate mechanism alone provides substantial gains. In contrast, removing the debate mechanism (\textsc{TraceRepair}$_{ND}$) results in 158 correct fixes, suggesting that trace-guided single-pass generation is slightly more effective than static debate alone. This highlights the strong role of runtime information in directly constraining the repair space.

\begin{table}[h]
  \vspace{-4mm}
  \caption{Ablation study on Defects4J v1.2 (SF bugs).}
  \label{tab:ablation_study}
  \footnotesize
  \centering
  \resizebox{\columnwidth}{!}{%
  \begin{tabular}{l l c c}
    \toprule
    \textbf{Variant} & \textbf{Configuration} & \textbf{Correct} & \textbf{Improv.} \\
    \midrule
    DeepSeek-V3.2 & Baseline & 102 & - \\
    \midrule
    \textsc{TraceRepair}$_{NT}$ & w/o Trace (Static Debate Only) & 154 & +50.9\% \\
    \textsc{TraceRepair}$_{ND}$ & w/o Debate (Trace Only)        & 158 & +54.9\% \\
    \textsc{TraceRepair}$_{NDS}$ & w/o Defensive Strategy         & 166 & +62.7\% \\
    \textsc{TraceRepair}$_{NCS}$ & w/o Causal Strategy            & 171 & +67.6\% \\
    \textsc{TraceRepair}$_{NSS}$ & w/o Semantic Strategy          & 169 & +65.7\% \\
    \midrule
    \textbf{\textsc{TraceRepair}$_{DSV3.2}$} & \textbf{Full Framework}     & \textbf{177} & \textbf{+73.5\%} \\
    \bottomrule
  \end{tabular}%
  }
  \vspace{-2mm}
\end{table}

Further analysis shows that all three debate strategies contribute to performance. Removing the Defensive strategy leads to the largest drop, decreasing from 177 to 166 fixes, followed by the Semantic strategy with 169 fixes and the Causal strategy with 171 fixes. The full framework achieves 177 correct fixes, representing a 73.5\% improvement over the baseline and consistently outperforming all ablated variants. These results indicate that runtime traces and the debate mechanism address complementary failure modes, and that the three debate strategies jointly contribute to more accurate and reliable repairs.




\vspace{0.5em}
\noindent\fbox{
  \parbox{0.95\linewidth}{
    \textbf{Summary of RQ2:} Ablation shows that runtime traces and debate independently improve results, and work best together. All three strategies are necessary, especially the Defensive one. Iterative debate fixes more bugs than a single generation attempt, though the benefits slow down in later rounds.
  }
}

\subsection{RQ3: Robustness and Generalization}
\label{sec:rq3}

LLM-based APR raises a fundamental question: whether models fix bugs by reasoning about program execution or by matching patterns from previously seen fixes. We study this issue from two aspects. First, we examine the reliability of the Probe Agent in instrumenting real-world programs. Second, we evaluate whether the framework can repair bugs that are not covered by the model’s training data.

\textbf{Reliability of Diagnostic Probing.}
Runtime traces are only effective when they can be reliably collected. If the Probe Agent generates instrumentation code that fails to compile, the trace-guided pipeline loses its evidential basis. To assess this risk, we measure the Instrumentation Success Rate across all 673 bugs in Defects4J, as reported in Table~\ref{tab:instrumentation_success}. \textsc{TraceRepair}$_{DSV3.2}$ achieves a success rate of 93.8\% under PFL, indicating strong robustness across diverse code contexts. \textsc{TraceRepair}$_{GPT3.5}$ attains 79.3\%, with failures more frequently occurring in complex methods. These results demonstrate that the Probe Agent can reliably produce valid instrumentation in most cases, ensuring that the downstream debate stage is consistently supported by executable runtime evidence.



\begin{table}[htbp]
    \centering
    \footnotesize
    \caption{Probe Agent instrumentation success rate.}
    \vspace{-3mm}
    \label{tab:instrumentation_success}
    \renewcommand{\arraystretch}{1.1} 
    \begin{tabular*}{\columnwidth}{@{\extracolsep{\fill}} l c c @{}}
        \toprule
        \textbf{Approach} & \textbf{PFL Success} & \textbf{MFL Success} \\
        \midrule
        \textsc{TraceRepair}$_{GPT3.5}$          & 534 / 673 (79.3\%) & 546 / 673 (81.1\%) \\
        \textsc{TraceRepair}$_{DSV3.2}$          & 631 / 673 (93.8\%) & 622 / 673 (92.4\%) \\
        \bottomrule
    \end{tabular*}

\vspace{-3mm}
\end{table}

{As shown in Table~\ref{tab:rq3_results}, we evaluate \textsc{TraceRepair} on \textsc{Recent-Java} using two backbone models, GPT-3.5 and DeepSeek-Coder. Under GPT-3.5, we compare against the vanilla model and ChatRepair, representing direct generation and iterative test-feedback repair, respectively. To examine whether the improvements of \textsc{TraceRepair} arise merely from access to runtime information, we additionally include two trace-based baselines. TracePrompt follows the trace-augmented prompting strategy proposed by Haque et al.~\cite{haque2025towards}, while InspectCoder performs LLM-guided runtime-state inspection through an interactive debugger~\cite{wang2026inspectcoder}.}

{All baselines are reproduced or adapted to the Java repair setting and rerun in our experimental environment using the same buggy programs, failure-triggering tests, backbone model, and validation protocol. We exclude retrieval-augmented methods such as REINFIX because they introduce external repair patterns at inference time and are therefore less directly comparable when evaluating repair performance on post-cutoff defects.}

\begin{table}[h]
  \caption{{Overall repair performance on the \textsc{Recent-Java} benchmark.}}
  \vspace{-3mm}
  \label{tab:rq3_results}
  
  \footnotesize
  \begin{tabular*}{\columnwidth}{@{\extracolsep{\fill}}llc@{}}
    \toprule
    \textbf{{Method}} & \textbf{{Backbone}} & \textbf{{\#Fixed}} \\
    \midrule
    {GPT-3.5}             & {GPT-3.5}        & {4} \\
    {ChatRepair}          & {GPT-3.5}        & {8} \\
    {TracePrompt}         & {GPT-3.5}        & {1} \\
    {InspectCoder}        & {GPT-3.5}        & {3} \\
    {\textsc{TraceRepair}}& {GPT-3.5}        & {\textbf{10}} \\
    \midrule
    {DeepSeek-Coder}             & {DeepSeek-Coder} & {2} \\
    {\textsc{TraceRepair}}& {DeepSeek-Coder} & {\textbf{6}} \\
    \bottomrule
  \end{tabular*}
  \vspace{-3mm}
\end{table}

{Under GPT-3.5, the vanilla baseline repairs 4 of the 21 defects, while ChatRepair increases this number to 8 through iterative test feedback. TracePrompt and InspectCoder repair 1 and 3 defects, respectively, indicating that providing runtime information alone does not necessarily lead to effective repair. In contrast, \textsc{TraceRepair} repairs 10 defects by using runtime observations not only as additional context, but also to support the generation, comparison, and refinement of competing repair hypotheses.}

With DeepSeek-Coder, \textsc{TraceRepair} increases the number of repaired defects from 2 to 6 compared with the corresponding vanilla baseline. These results indicate that the improvements of \textsc{TraceRepair} are not limited to a particular backbone model. They also suggest that its effectiveness depends on how runtime observations are incorporated into the repair process, rather than merely on whether such information is provided to the model.

\vspace{0.5em}
\noindent\fbox{
  \parbox{0.95\linewidth}{
    \textbf{Summary of RQ3:} Even on defects that emerged after the model's training cutoff, \textsc{TraceRepair} consistently outperforms both static and conversational baselines, demonstrating that its repair capability stems from dynamic reasoning over execution evidence rather than memorization of training data.
  }
}

\subsection{RQ4: Efficiency and Cost}
\label{sec:rq4}
Repair accuracy alone does not determine whether a framework is practical at scale. Methods that incur excessive token usage or long execution time are difficult to deploy in real-world development workflows. To evaluate practicality, we measure three metrics for each approach: average token consumption, monetary cost, and end-to-end execution time per bug. Table~\ref{tab:cost_efficiency} compares \textsc{TraceRepair} with prior methods that report efficiency-related metrics, including ChatRepair, RepairAgent, and AdverIntent-Agent. We base our comparison on the reported results of these methods to ensure consistency with their original evaluation settings.


\begin{table}[htbp]
  \caption{Efficiency and cost comparison (average per bug).}
  \vspace{-2mm}
  \label{tab:cost_efficiency}
  \footnotesize
  \centering
  \renewcommand{\arraystretch}{1.2}
  \begin{tabular*}{\columnwidth}{@{\extracolsep{\fill}} l r c c @{}} 
    \toprule
    \textbf{Approach} & \textbf{Tokens} & \textbf{Cost} & \textbf{Time} \\
    \midrule
    ChatRepair & 467k & \$0.42 & - \\
    AdverIntent-Agent & 438k & - & - \\
    RepairAgent  & 270k & \$0.14 & 920.0s \\
    \midrule
    \textbf{\textsc{TraceRepair}$_{GPT3.5}$} & \textbf{85k} & \textbf{\$0.11} & \textbf{251.7s} \\
    \textbf{\textsc{TraceRepair}$_{DSV3.2}$} & \textbf{77k} & \textbf{\$0.014} & 472.9s \\
    \bottomrule
  \end{tabular*}
  \vspace{-2mm}
\end{table}

\textbf{Token and Cost Analysis.}
Conversational repair approaches such as ChatRepair rely on multi-turn interactions to reason about program behavior, which leads to increasing token usage as context accumulates across iterations. In contrast, \textsc{TraceRepair} reduces this overhead by providing repair agents with concrete runtime evidence prior to patch generation, thereby limiting the need for extended dialogue. This difference is reflected in the efficiency metrics. As shown
in Table~\ref{tab:cost_efficiency}, \textsc{TraceRepair}$_{GPT3.5}$ consumes 85k tokens per bug on average, compared to 467k for ChatRepair and 270k for RepairAgent. \textsc{TraceRepair}$_{DSV3.2}$ further reduces this to 77k tokens. In terms of cost, \textsc{TraceRepair}$_{DSV3.2}$ averages 0.014 dollars per bug, and \textsc{TraceRepair}$_{GPT3.5}$ 0.11 dollars, both substantially lower than ChatRepair at 0.42 dollars. The cost efficiency of \textsc{TraceRepair}$_{DSV3.2}$ is attributed to both reduced token consumption and the lower pricing of the underlying model, making it particularly suitable for large-scale deployment.

{Table~\ref{tab:token_breakdown} reports the agent-level token usage of \textsc{TraceRepair}. For \textsc{TraceRepair}$_{GPT3.5}$, the Probe Agent uses 4.2k tokens per bug, accounting for 4.9\% of the total. For \textsc{TraceRepair}$_{DSV3.2}$, it uses 5.2k tokens, accounting for 6.7\%. In both settings, more than 90\% of the tokens are consumed by the Repair Agents, while the Judge Agent accounts for only 2.2\%. Thus, most LLM token consumption comes from candidate generation, critique, and refinement, whereas diagnostic probing and final arbitration account for only a small share.}

\begin{table}[t]
\vspace{-2mm}
\footnotesize
\caption{{Token breakdown of \textsc{TraceRepair} by agent (average per bug).}}
\label{tab:token_breakdown}
\small
\begin{tabular}{lcccc}
\toprule
{Variant} & {Total} & {Probe} & {Repair} & {Judge} \\
\midrule
{\textsc{TraceRepair}$_{GPT3.5}$}
& {85k} & {4.2k (4.9\%)} & {78.9k (92.9\%)} & {1.9k (2.2\%)} \\
{\textsc{TraceRepair}$_{DSV3.2}$}
& {77k} & {5.2k (6.7\%)} & {70.1k (91.1\%)} & {1.7k (2.2\%)} \\
\bottomrule
\end{tabular}
\vspace{-2mm}
\end{table}


\textbf{Runtime Analysis.}
Execution time is influenced by factors beyond the repair framework itself, including API latency and runtime environment, and should therefore be interpreted as indicative rather than definitive. Under this context, the comparison with RepairAgent remains informative: RepairAgent requires 920.0s per bug on average, while \textsc{TraceRepair}$_{GPT3.5}$ and \textsc{TraceRepair}$_{DSV3.2}$ reduce this to 251.7s and 472.9s, respectively. The higher latency of the DeepSeek-V3.2 variant is primarily due to differences in API response characteristics rather than the repair process. A breakdown by repair outcome provides additional insight. For successfully fixed bugs, \textsc{TraceRepair}$_{DSV3.2}$ completes in 259.3s on average, while \textsc{TraceRepair}$_{GPT3.5}$ requires 119.8s. In contrast, unsuccessful runs are significantly longer, reaching up to 898.3s for DeepSeek-V3.2, as the framework exhausts its iteration budget. This disparity suggests that when runtime traces provide sufficient diagnostic signal, the repair process converges quickly, whereas prolonged runs typically indicate that the available execution evidence is insufficient to resolve the defect.



\vspace{0.5em}
\noindent\fbox{
  \parbox{0.95\linewidth}{
    \textbf{Summary of RQ4:} \textsc{TraceRepair} consumes far fewer tokens and incurs lower cost than all baselines, as trace guided reasoning replaces repeated rounds of dialogue. When traces provide sufficient evidence, repair sessions converge quickly; sessions without a fix use up the full iteration budget before terminating.
  }
}

\section{Threats to Validity}
\label{sec:threats}

We identify three categories of threats to the validity of our findings.

\textbf{Internal Validity.}
A key concern in LLM-based APR is data leakage, where models may recall training data rather than reason about bugs. We address this issue through the \textsc{Recent-Java} benchmark, which contains defects after the training cutoff of our backbone models. The successful repair of 10 unseen bugs (Section~\ref{sec:rq3}) and independent manual verification of generated patches by two authors suggest that the performance is not due to memorization.

\textbf{External Validity.}
The diagnostic instrumentation strategy of \textsc{TraceRepair} is built around runtime observation rather than any particular language feature, which positions it well for adaptation beyond Java. We selected Java as the primary evaluation language because Defects4J is the most mature and widely adopted benchmark in the APR community, enabling direct comparison with prior work. Extending the Probe Agent to other languages remains a natural direction for future work.

\textbf{Construct Validity.}
\textsc{TraceRepair} depends on test suites that exercise the buggy logic sufficiently to produce a trace. Under weak test coverage or flaky test conditions, the Probe Agent may fail to capture relevant runtime state, leaving the debate agents without grounded evidence. Our experiments with coarse-grained fault localization suggest that the multi-agent debate mechanism retains reasonable resilience under noisy trace data, though improving robustness under low coverage conditions remains future work.

\section{Related Work}
\label{sec:related_work}

Related work on APR spans heuristic search, neural and LLM-based generation, trace-based repair, and agentic debugging. \textsc{TraceRepair} bridges trace-based repair and agentic reasoning.

\textbf{From Heuristics to Neural Translation.}
Traditional APR methods generate patches through search, constraints, or repair templates. GenProg~\cite{le2011genprog} and ARJA~\cite{yuan2018arja} use genetic search but may produce test-overfitting patches~\cite{qi2015analysis}; Angelix~\cite{mechtaev2016angelix} and S3~\cite{le2017s3} rely on symbolic constraints; and TBar~\cite{liu2019tbar} and FixMiner~\cite{koyuncu2020fixminer} apply predefined fix patterns. Neural approaches such as CoCoNut~\cite{lutellier2020coconut}, CURE~\cite{jiang2021cure}, and Recoder~\cite{zhu2021syntax} learn mappings from buggy to fixed code, but remain limited by their training distributions and struggle with unseen logic and project-specific identifiers.

\textbf{Generative Repair and Retrieval Augmentation.}
Large Language Models broadened APR by enabling repair without prior exposure to similar bugs. Models such as Codex~\cite{chen2021evaluating}, StarCoder~\cite{li2023starcoder}, and DeepSeek-Coder~\cite{guo2024deepseek} provide strong code reasoning capabilities, while approaches such as AlphaRepair~\cite{xia2023automated} and GAMMA~\cite{zhang2023gamma} exploit LLMs through cloze-style generation. To mitigate hallucination in pure generation~\cite{ji2023survey}, retrieval-augmented frameworks such as REINFIX~\cite{zhang2025repair} and CEDAR~\cite{nashid2023retrieval} provide relevant fix patterns. However, such methods are less effective for novel state-dependent faults without similar historical fixes.

{\textbf{Trace-Based Program Repair.}
Existing studies have used runtime information to support program repair and debugging. SelfAPR incorporates test execution diagnostics into neural patch generation to capture project-specific failure information~\cite{ye2022selfapr}. Haque et al. augment LLM repair prompts with execution traces and show that raw traces require suitable processing to be effective~\cite{haque2025towards}. TraceFixer uses execution traces for program repair~\cite{bouzenia2023tracefixer}, while InspectCoder enables LLMs to inspect runtime states through an interactive debugger~\cite{wang2026inspectcoder}. Zhong et al. and Kang et al. further use runtime verification and scientific debugging to support failure diagnosis~\cite{zhong2024debug,kang2025explainable}. Unlike these approaches, \textsc{TraceRepair} uses runtime observations throughout the repair process to generate, compare, and refine competing repair hypotheses through multi-agent debate.}

\textbf{Agentic Debugging and Reasoning.}
Recent work has moved toward autonomous agents that interact with the codebase rather than passively generating patches. Self-Debugging~\cite{chen2024teaching} and ChatRepair~\cite{xia2024automated} use conversational feedback loops, while RepairAgent~\cite{bouzenia2025repairagent} and SWE-agent~\cite{yang2024swe} further incorporate tool use for repository exploration and analysis. Benchmarks such as SWE-Bench~\cite{jimenez2024swe} evaluate these agents on repository-level issue resolution. In contrast, this work focuses on the repair stage where faulty locations and relevant tests are available, and explores how runtime traces can provide fine-grained behavioral evidence to guide patch generation. Extending trace-guided repair to repository-level issue resolution remains an important direction for future work.

\section{Conclusion}
\label{sec:conclusion}

This paper presents \textsc{TraceRepair}, a multi-agent APR framework that incorporates runtime trace evidence into the patch generation process. \textsc{TraceRepair} produces more accurate and reliable fixes.

Extensive experiments show that runtime traces consistently improve fault localization and patch correctness, while the multi-agent debate mechanism reduces hallucination by anchoring candidates to observed runtime facts. Evaluation on unseen defects confirms that the framework's effectiveness stems from dynamic reasoning rather than memorization. We hope this work offers a useful perspective on combining dynamic analysis with LLM-based repair, and encourages further exploration of execution-driven approaches.


\begin{acks}
This research was supported by the National Natural Science Foundation of China (Grant Nos.61902295, 62141220 and 62372376). Bo Shen is the corresponding author.
\end{acks}

\section*{Data Availability}
We provide a replication package containing the implementation of \textsc{TraceRepair}, the scripts, and all data needed to reproduce our results. The package is anonymously available at: \url{https://doi.org/10.5281/zenodo.19252356}.

\balance
\bibliographystyle{ACM-Reference-Format}
\bibliography{sample-base}

\end{document}